\documentclass[12pt]{iopart}

\usepackage[]{graphicx}

\begin{document}
\JPA

\letter {A variational principle behind van der Waals picture of strongly coupled plasmas}

\author {Hiroshi Frusawa\footnote[1]{e-mail:
frusawa.hiroshi@kochi-tech.ac.jp}}

\address{Soft Matter Laboratory, Kochi University of Technology, Tosa-Yamada, Kochi 782-8502, Japan}

\begin{abstract}
Various strong coupling theories of the one-component plasma have successfully predicted the thermodynamic and structural properties by separating the Coulomb potential into short- and long-ranged parts in {\itshape ad hoc} ways. Moreover, it has been demonstrated that the density-density correlation function in a mimic system with only the short-ranged interactions resembles that of the full Coulomb system, revealing that the van der Waals picture applies to the strongly coupled Coulomb systems. Here we provide a variational theory forming the basis of the van der Waals picture. Our approach provides hybrid formulations which combine both the liquid state theory and statistical field theory; essential use is made of the coarse-grained system with only long-raged part of Coulomb interactions as a reference system in introducing both lower bound variational principle and strong coupling expansion.
\end{abstract}

\section{Introduction}

The van der Waals picture 
\cite{widom, wca, hansen} 
is a robust framework for liquids with nontrivial density correlations.
According to this treatment, the harshly repulsive interactions at short distances determine the average relative arrangements of components, and slowly varying long-ranged interactions play a minor role in the liquid structure.
One of well-known prescriptions for the potential separation is the systematic Weeks-Chandler-Andersen (WCA) method
\cite{wca, hansen} 
which has been successful for the interaction potentials with minima such as the Lennard-Jones potential.

The WCA separation, however, is not applicable to the Coulomb interaction potential with neither minimum nor a characteristic length. A typical potential form due to the Coulomb interaction between one component charged particles
is given by $\gamma v(r)=\gamma(a/r)$, where $|{\bf r}|=r$, the coupling constant $\gamma=\beta(ze)^2/(4\pi\epsilon\,a)$ is defined by the inverse thermal energy $\beta$, the charge per particle $ze$, the dielectric permittivity $\epsilon$, and a reference length $a$ equal to the Wigner-Seitz radius \cite{review}: $a=(3/4\pi\rho)^{1/3}$ where the concentration of charged particles is given by $\rho=N/V$ with $N$ being the total number of particles and $V$ the system volume. A prototype of the Coulomb systems is the one-component plasma (OCP) \cite{review}, a uniform system of point ions in a neutralizing background, which we will consider here.
This model can in fact be regarded as a limiting case of real matter such as liquid metals, salt-free colloidal dispersions, and so on. The long-range Coulombic nature  of the OCP in the strong coupling (SC) regime satisfying that $\gamma>>1$ leads to the Wigner crystallization with large lattice constant comparable to the wavelength of visible light, which is an underlying cause of photonic crystal formation observed for salt-free colloidal dispersions \cite{warren}.

When we separate the Coulomb potential $\gamma v$ into short- and long-ranged parts, $\gamma v_S$ and $\gamma v_L$, with $v_L$ supposed to have a finite value at zero separation, we have had a choice, as short-ranged contribution $\gamma v_S$, to use a hard core potential with an effective hard-core diameter, an adjustable parameter in the SC regime \cite{review}. In addition to the hard-core type division, there have been other various ad-hoc methods \cite{review, original, weeks_ocp, totsuji, onsager, rosenfeld_lb, brilliantov, frusawa_lb} of the potential separation. In the random phase approximation (RPA), the previous SC theories with various optimized forms of the long-ranged potentials $\gamma v_L$ have provided the internal correlation energy in excellent agreement with simulation results over a wide range of coupling constant $\gamma$ \cite{review, original, weeks_ocp, totsuji, onsager, rosenfeld_lb, brilliantov}.
Moreover, it has been recently demonstrated by use of local molecular field (LMF) theory \cite{weeks_ocp} that the structures of the OCP are related to those of a mimic system with only short-ranged interactions in an external field incorporating a mean-field average of the remaining long-ranged interaction system: the density-density correlation function in the mimic system resembles that of the full Coulomb system.
The successful predictions of previous SC theories for both thermodynamic and structural properties
implies the efficiency of the Coulomb potential separation along the van der Waals picture; however these treatments have not clarified both a criterion for the division of Coulomb potential and a functional-integral form validating the RPA in the SC regime.

In order to verify the van der Waals picture of the strongly coupled OCP, we must thus face the following fundamental questions: (i) What theory forms the basis of separating into two parts the Coulomb potential without characteristic length scales? (ii) Why is the RPA (or the saddle-point approximation), validated so far in the weak coupling regime \cite{samuel, brown, netz_strong}, relevant even to the strongly coupled systems? Our main aim is to provide a solution to these problems by developing a hybrid theory which combines both the liquid state theory and the statistical field theory; essential use is made of both lower bound variational principle, other than the well-known upper bound approach \cite{hansen, frusawa_lb} referred to as the thermodynamic perturbation theory, and strong coupling expansion around saddle-point which can actually be formulated via the standard field-theoretic method, contrary to the statements in the literature \cite{samuel, brown, netz_strong}. In the following, we first establish a variational principle to optimize the potential separation by maximizing the lower bound of the free energy.
Next we present a functional integral form to which the RPA is applicable in the SC regime. Our variational principle offers not only a framework for addressing the above fundamental problems, but also an alternative to deriving the lower bound of internal correlation energy evaluated so far in various ways \cite{review, totsuji, onsager, rosenfeld_lb, brilliantov, frusawa_lb, lieb}. Finally we compare our result of potential separation with that of the LMF theory 
\cite{weeks_ocp}. 

\section{Outline of the lower bound approach}

To answer the first question described in the introduction, we start with not the upper bound due to the Gibbs--Bogoliubov inequality 
\cite{hansen, frusawa_lb}, 
but the lower bound \cite{hansen} of the true free energy $F\{v\}$ for the OCP with full Coulomb interactions. Taking the long-ranged potential $v_L$ as a reference system potential in the Coulomb separation of $\gamma v=\gamma v_S+\gamma v_L$, we have the inequality for the lower bound $\mathcal{L}\{v_L;h\}$ as a functional of both $v_L$ (or $v_S=v-v_L$) and the pair correlation function $h(r)$ of {\itshape not a reference but a full Coulomb system} \cite{hansen}:
   \begin{equation}
   \frac{\mathcal{L}\{v_L;h\}}{N}=
   \frac{\beta F\{v_L\}}{N}+
   \frac{\gamma\rho}{2}\int_{{\bf r}}h(r)v_S(r)\leq\frac{\beta F\{v\}}{N},
   \label{start}
   \end{equation}
where $F\{v_L\}$ is the free energy of a long-ranged interaction system, $\int_{{\bf r}}$ denotes the integral $\int d{\bf r}$.
Evaluating the free energy $F\{v_L\}$ in the RPA, it will be shown that {\itshape maximization} of the lower bound functional $\mathcal{L}$ with respect to $v_L$ yields
   \begin{equation}
   \fl\qquad\qquad
   \frac{\mathcal{L}_{\mathrm{max}}}{N}\equiv
   \frac{\mathcal{L}\{v_L^*;h\}}{N}=
   \frac{\beta F\{-c/\gamma\}}{N}
   +\frac{\rho}{2}\int_{{\bf r}}h(r)\left[\,
   \gamma v(r)+c(r)
   \,\right].
   \label{max}
   \end{equation}
This means that the optimum potentials, $v_L^*$ and $v_S^*$, are
   \begin{equation}
   \gamma v^*_L=-c;\qquad \gamma v^*_S=\gamma v + c=c_{sr},
   \label{optimum}
   \end{equation}
where $c$ denotes the direct correlation function and $c_{sr}$ corresponds to the so-called short-ranged direct correlation function \cite{evans}. In other words, the long-range part $\gamma v^*_L$ of Coulomb potential satisfies the Orstein-Zernike (OZ) integral equation \cite{hansen}:
   \begin{equation}
   h(r)=-\gamma v^*_L(r)-\rho\int_{{\bf r}'}
   \gamma v_L^*(|{\bf r}-{\bf r}'|)h(r').
   \label{oz}
   \end{equation}
Using diagrammatic analysis, the exact closure to the OZ equation is expressed as
   \begin{equation}
   g(r)=\exp\left[
   -\gamma v(r)+\gamma v_L(r)+h(r)-b(r)
   \right],
   \label{closure}
   \end{equation}
where the radial distribution function $g(r)$ is defined by $g=1+h$ and the bridge function $b(r)$ denotes the negative of the sum of elementary diagrams \cite{hansen}.

The relations (\ref{start}) to (\ref{optimum}) state that the variational approach for the lower bound functional $\mathcal{L}\{v_L;h\}$ provides a first principle for separating the Coulomb potential. A question still remains whether the real density correlations represented by $h(r)$ are mimicked by the short-ranged system with the interaction potential $c_{sr}$. Let $F_{sr}=F\{v\}-F\{-c/\gamma\}$ and $F\{c_{sr}/\gamma\}$, respectively, the short-ranged contribution to the full free energy and the mimic free energy associated with short-ranged system. In the RPA of the long-ranged contribution, the mimic free energy $F\{c_{sr}/\gamma\}$ is identified with $F_{sr}$, giving
   \begin{equation}
   \frac{\beta F\{c_{sr}/\gamma\}}{N}=\frac{\beta F_{sr}}{N}\approx \frac{\rho}{2}\int_{{\bf r}}h(r)c_{sr}(r)
   \label{mimic}
   \end{equation}
in the approximation, $\beta F_{sr}\approx\mathcal{L}_{\mathrm{max}}-\beta F\{v_L\}$. Equation (\ref{mimic}) leads to
   \begin{equation}
   \frac{2}{N}
   \frac{\delta \beta F\{c_{sr}/\gamma\}}{\delta c_{sr}}
   =\rho h(r),
   \label{correlation_ocp}
   \end{equation}
indicating that the mimic system with only short-ranged interaction potential, $\gamma v^*_S$, can reproduce the pair correlation function $h(r)$ of full Coulomb system in line with the van der Waals picture.


\section{Strong Coupling Approximation of the OCP}

All of the discussions made in the previous section have relied on the RPA, though the RPA has not been verified in the SC regime: while previous field theories have formulated the RPA only in the weak coupling regime $\gamma<<1$ 
\cite{samuel, brown, netz_strong}, 
strong coupling theories have successfully used the RPA without the field-theoretic justification
\cite{review, totsuji, onsager, rosenfeld_lb, brilliantov}. 
We then reveal below the underlying functional integral form behind the RPA in the SC regime.

For later convenience, we introduce the free energy functional $\mathcal{A}\{v_L\}$ including the self-energy via the expression:
   \begin{equation}
   \fl\qquad\qquad
   \beta F\{v_L\}=\beta\mathcal{A}\{v_L\}-\frac{\gamma v_L(0)}{2}
   \ln\left<
   e^n\right>
   =\beta\mathcal{A}\{v_L\}-
   \frac{\gamma v_L(0)}{2}N,
   \label{def_A}
   \end{equation}
with the bracket denoting the average as follows: $<\mathcal{O}>=\mathrm{Tr}\,\mathcal{O}e^{-\gamma U\{v_L\}}/\mathrm{Tr}\,e^{-\gamma U\{v_L\}}$, where $\mathrm{Tr}\equiv\sum_{n=0}^{\infty}(z^n/n!)\,\int_{{\bf r}_1\cdots{\bf r}_n}$, $z$ the activity, and the total electrostatic interaction energy $U\{v_L\}$ includes the self-energy of ions. In the last equality in eq. (\ref{def_A}), use has been made of the relation: $\left<\e^n\right>=e^{\left<n\right>}=e^N$. The interaction energy $\gamma U\{v_L\}$ is expressed as
   \begin{eqnarray}
   \gamma U\{v_L\}=\frac{\gamma}{2}\int_{{\bf r}_1{\bf r}_2}
   \delta\hat{\rho}_1\,v_L(r_{12})\,\delta\hat{\rho}_2
   \label{interaction},
   \end{eqnarray}
setting for abbreviation that $\delta\hat{\rho}({\bf r})=\hat{\rho}({\bf r})-\rho$, $\hat{\rho}({\bf r})=\sum_i\delta({\bf r}-{\bf r}_i)$, $\delta\hat{\rho}_1=\delta\hat{\rho}({\bf r}_1)$ and $r_{12}=|{\bf r}_1-{\bf r}_2|$. Along the Hubbard-Stratonovich transformation 
\cite{samuel, brown, netz_strong}, we rewrite eq. (\ref{interaction}) as
   \begin{eqnarray}
   e^{-\gamma U\{v_L\}}&=&\frac{1}{\mathcal{N}}\int_{\phi}e^{-\gamma S_0\{\phi\}+\gamma\int_{{\bf r}}i\phi({\bf r})\left\{\hat{\rho}({\bf r})-\rho\right\}}\nonumber\\
   S_0\{\phi\}&=&\frac{1}{2}\int_{{\bf r}_1{\bf r}_2}
   \phi_1\,
   v_L^{-1}(r_{12})\,
   \phi_2
   \nonumber
   \end{eqnarray}
with $\mathcal{N}=\int_{\phi}e^{-S_0\{\phi\}}$. Consequently, we obtain the functional integral form of $\mathcal{A}\{v_L\}$:
   \begin{eqnarray}
   &&e^{-\beta \mathcal{A}\{v_L\}}=\frac{1}{\mathcal{N}}\int_{\phi}\exp\left(-\gamma S\{\phi\}\right)\nonumber\\
   &&\gamma S\{\phi\}=\gamma S_0\{\phi\}+\gamma\rho\int_{{\bf r}}i\phi({\bf r})
   -z\int_{{\bf r}}e^{\gamma i\phi({\bf r})}.
   \label{fieldstart}
   \end{eqnarray}
Since the action $\gamma S\{\phi\}$ is proportional to the coupling constant $\gamma$, our functional integral form verifies that the perturbation expansion around the saddle-point satisfying that $\delta S/\delta\phi|_{\phi=i\phi^*}=0$ is allowed to use in the SC regime.

Before proceeding, let us compare the present action $\gamma S\{\phi\}$ with the previous one for the weak coupling regime \cite{netz_strong}: $\gamma^{-1}S'\{\psi\}=\gamma^{-1}S_0\{\psi\}-\rho\int_{{\bf r}}i\psi({\bf r})-ze^{\gamma v_L(0)/2}\int_{{\bf r}}e^{-i\psi({\bf r})}$ where $v_L$ should be replaced by $v$ and elimination of the self-energy is considered in the last exponent on the right hand side
\cite{netz_strong} other than our treatment in eq. (\ref{def_A}).
The comparison indicates that our statistical field theory has introduced a normalized field $\phi=-\psi/\gamma$ instead of the standard potential field $\psi$. Our selection of potential variable is indispensable for picking up inherent fluctuations from $\psi$ whose variations multiplied by $\gamma$ are amplified extremely in the SC regime.

The saddle-point equation, given by $\delta S/\delta\phi|_{\phi=i\phi^*}=0$, reads
   \begin{equation}
   \phi^*({\bf r}_1)=\int_{{\bf r}_2}v^*_L(r_{12})
   \left[
   ze^{-\gamma\phi^*({\bf r}_2)}-\rho
   \right].
   \label{mf_uniform}
   \end{equation}
The prefactor $\gamma$ appearing in the exponent leads to the exact solution in the SC limit: we obtain that $\lim_{\gamma\rightarrow\infty}\phi^*\equiv 0$ and $z=N/V=\rho$; otherwise there are no self-consistent solutions, because the above term, $e^{-\gamma\phi^*}$, goes to zero for $\phi>0$ and diverges for $\phi<0$ in the limit $\gamma\rightarrow\infty$. While eq. (\ref{mf_uniform}) holds in the SC regime $\gamma>>1$, the expression is identified with the Poisson-Boltzmann equation valid in the weak coupling regime $\gamma<<1$ \cite{netz_strong} if $\phi^*$ and $v_L^*$ are replaced by $\psi/\gamma$ and the full potential $v$, respectively. The availability of the saddle-point equation in both the strong and weak coupling regimes seems to be one of the reasons for the success so far of the RPA, or the saddle-point approximation, in the intermediate regime \cite{review, original, weeks_ocp, totsuji, onsager, rosenfeld_lb, brilliantov}.

Let us write $\phi=i\phi^*+\chi/\gamma$ to expand $S\{\phi\}$ around the saddle point $\phi^*=0$. In the RPA, we find from eqs. (\ref{def_A}) to (\ref{fieldstart}) that the long-ranged contribution $F\{v_L\}$ to the free energy is of the form:
   \begin{eqnarray}
   \beta F\{v_L\}=\gamma S\{i\phi^*=0\}-\ln
   \frac{1}{\mathcal{N}}
   \int_{\chi}
   \exp\left[-\Delta S\{\chi\}\right]-\frac{N\gamma}{2}v(0),
   \label{rpa}
   \end{eqnarray}
where the first term $\gamma S\{0\}$ on the right hand side corresponds to the mean-field free energy, and the quadratic action $\Delta S\{\chi\}$ in the second term is associated with the expansion for the last term on the right hand side of eq. (\ref{fieldstart}) such that $e^{\gamma i\phi}=e^{-\gamma\phi^*+i\chi}=e^{-\gamma\phi^*}(1+i\chi-\chi^2/2)$, giving
   \begin{equation}
   \Delta S\{\chi\}=\frac{1}{2\gamma}
   \int_{{\bf r}_1{\bf r}_2}\chi_1\left[v_L^{-1}(r_{12})+\gamma\rho \delta(r_{12})\right]\chi_2.
   \label{quadratic}
   \end{equation}
Combining eqs. (\ref{start}), (\ref{rpa}) and (\ref{quadratic}), we obtain
   \begin{equation}
   \fl\quad
   \frac{\mathcal{L}\{v_L;h\}}{N}
   =-\frac{\gamma v_L(0)}{2}-\frac{\rho\gamma}{2}\int_{{\bf r}}v_S(r)
   +\frac{\rho\gamma}{2}\int_{{\bf r}}g(r)v_S(r)
   +\frac{1}{2\rho}\int_{{\bf k}}\ln\left[1+\rho\gamma \tilde{v}_L(k)\right],
   \label{lower_functional}
   \end{equation}
where we have used the relation $h(r)=-1+g(r)$ and $\tilde{v}_L(k)$ denotes the Fourier transform of the interaction potential $v_L$. The maximum condition on the lower bound functional $\mathcal{L}$ is the functional differentiation of $\mathcal{L}\{v_L;h\}$ given by eq. (\ref{lower_functional}) with respect to $v_L$:
    \begin{equation}
    \left.
    \frac{\delta\mathcal{L}}{\delta v_L}\right|_{v_L=v_L^*}=0,
    \label{stationary}
    \end{equation}
yielding the OZ equation (\ref{oz}) within the RPA \cite{rosen_oz}. When we consider higher order terms beyond the RPA, we can evaluate from eq. (\ref{stationary}) correction potential to minus the direct correlation function $\gamma v_L^*=-c$ of long-ranged part optimized within the RPA. To be noted also, the second functional derivative of $\mathcal{L}$, arising from the logarithmic term on the right hand side of eq. (\ref{lower_functional}), has a negative sign, which proves that the maximum lower bound is located at the stationary point satisfying eq. (\ref{stationary}).

A set of our self-consistent equations is now complete by combining the OZ equation (\ref{oz}) (or eq. (\ref{stationary}) in the RPA), closure relation (\ref{closure}) and the saddle-point equation (\ref{mf_uniform}), which are the three simultaneous equations with three parameters: $\phi^*({\bf r})$, $v_L(r)$ and $h(r)$.

\section{An alternative to the lower bound of internal correlation energy}

Let us relate the best lower bound of internal correlation energy $e_L$ per particle in the SC limit to the optimized lower bound functional $\mathcal{L}_{max}$ of the free energy via the inequality:
   \begin{equation}
   e_L\leq\lim_{\gamma\rightarrow\infty}\frac{\gamma}{N}\left(
   \frac{\partial\mathcal{L}_{max}}{\partial\gamma}
   \right).
   \label{to_energy}
   \end{equation}
We demonstrate below that eq. (\ref{to_energy}) provides an alternative approach to the lower bound obtained so far \cite{review, totsuji, onsager, rosenfeld_lb, brilliantov, frusawa_lb, lieb}. Since the last term on the right hand side of eq. (\ref{lower_functional}) yields a negligible term in the SC limit, we obtain
   \begin{equation}
   \fl\qquad\quad
   \lim_{\gamma\rightarrow\infty}\frac{\gamma}{N}\left(
   \frac{\partial\mathcal{L}_{max}}{\partial\gamma}\right)
   =\frac{c(0)}{2}-\frac{\rho}{2}\int_{{\bf r}}\left[\,
   \gamma v(r)+c(r)\,\right]+\frac{\rho}{2}\int_{{\bf r}}g(r)\left[\,
   \gamma v(r)+c(r)\,\right]
   \label{limit_functional}
   \end{equation}
where use has been made of the relation $\gamma(\partial c(r)/\partial \gamma)=c(r)$ following the approximate solutions known in the liquid state theory \cite{onsager, rosenfeld_lb, evans}. The relations (\ref{to_energy}) and (\ref{limit_functional}) with the inequality $g(r)\geq 0$ lead to
   \begin{equation}
   e_L=\frac{c(0)}{2}-\frac{\rho}{2}\int_{{\bf r}}\left[\,
   \gamma v(r)+c(r)\,\right],
   \label{lower_energy}
   \end{equation}
which is in agreement with that derived in terms of the liquid state theory \cite{onsager, rosenfeld_lb, evans}. In either the soft mean spherical approximation (SMSA) or the hypernetted--chain (HNC) approximation of the closure relation (\ref{closure}), eq. (\ref{lower_energy}) has been found to give the lower bound \cite{onsager, rosenfeld_lb, evans} which is identical or quite similar to the Lieb-Narnhofer bound \cite{review, onsager, frusawa_lb, lieb} obtained from the ionic sphere model, or the Onsager ball model.

We would like to also note the connection between the lower bound functional and the correlation functional of free energy derived in the liquid state theory \cite{morita}. It is easy to check that the lower bound functional in the RPA is identified with the correlation functional in the SMSA. Since we have that $\int_{{\bf r}}g(r)[\gamma v(r)+c(r)]=0$ in the SMSA, it follows from eq. (\ref{lower_functional}) that
   \begin{equation}
   \frac{\mathcal{L}_{max}}{N}=\frac{c(0)}{2}-\frac{\rho}{2}\int_{{\bf r}}
   \left[\gamma v(r)+c(r)\right]
   +\frac{1}{2\rho}\int_{{\bf k}}\ln\left[1-\rho\tilde{c}(k)\right]
   \end{equation}
which is nothing but the SMSA functional \cite{onsager, rosenfeld_lb}. In the HNC approximation, on the other hand, the bridge function $b(r)$ in eq. (\ref{closure}) is ignored, reducing to the closure $\gamma v+c=c_{sr}=h-\ln g$. In the quadratic expansion of $\ln g\approx h-(h^2/2)$ valid for $|h|<<1$, the approximate HNC equation reads $\gamma v+c=c_{sr}\approx h^2/2$. We plug the approximate equation only into the integral $\int_{{\bf r}}g(r)[\gamma v(r)+c(r)]$ while leaving the integral $\int_{{\bf r}}\gamma v(r)+c(r)=\int_{{\bf r}}h(r)-\ln g(r)$, because the effective exclusion zone satisfying that $g(r)=0$ (or $h(r)=-1$), the main contribution to the latter integration, is irrelevant to the former. We thus have that $\int_{{\bf r}}g(r)[\gamma v(r)+c(r)]\approx(1/2)\int_{{\bf r}}h^2(r)+\mathcal{O}[h^3]$, arriving at the HNC functional form \cite{rosenfeld_lb}:
   \begin{equation}
   \fl\qquad\quad
   \frac{\mathcal{L}_{max}}{N}\approx\frac{c(0)}{2}-\frac{\rho}{2}\int_{{\bf r}}
   \left[\gamma v(r)+c(r)\right]
   +\frac{\rho}{4}\int_{{\bf r}}h^2(r)
   +\frac{1}{2\rho}\int_{{\bf k}}\ln\left[1-\rho\tilde{c}(k)\right].
   \end{equation}
Thus, we have seen that the present formulations based on the lower bound variational principle are consistent with previous forms of the conventional liquid state theory in the whole coupling range.

\section{Comparison with the LMF theory}

The LMF theory constitutes the potential separation, $v=v_S+v_L$, and the self-consistent LMF equation for the effective external field incorporating the long-ranged part \cite{weeks_ocp}. 
The LMF simulations
\cite{weeks_ocp} 
solve the LMF equation combined with Monte Carlo simulations of the short-ranged interaction system in the external field obtained from the LMF equation.
The mean-field equation (\ref{mf_uniform}) obtained here coincide with the LMF equation in the low density approximation named mimic Poisson-Boltzmann approximation \cite{weeks_ocp} when we change the variable as $\phi^*\rightarrow\psi/\gamma$.

        \begin{figure}[hbtp]
        \begin{center}
	\includegraphics[bb=50 0 612 430,clip,width=8cm]{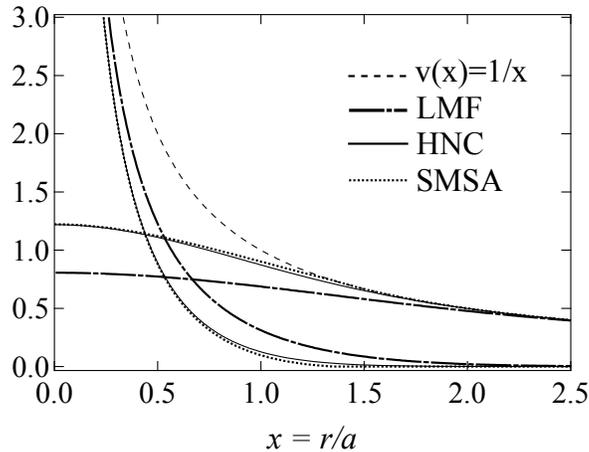}
	\end{center}
	\caption{Separation of the full Coulomb potential $v(x)=1/x=a/r$ at $\gamma=140$ in three ways. While the long-raged potentials converge to the full potential, the short-ranged remainders go to zero within the present x-range. Our separations in two approximations, HNC approximation and SMSA, are close to each other but deviated from the LMF division.}
       \end{figure}
       
Figure 1 compares the potential separations of the OCP at $\gamma=140$: the LMF division that $v_L^*(r)=\mathrm{erf}(x/1.4)/x$  with $x=r/a$ 
\cite{weeks_ocp} 
and ours taking two typical approximate solutions of $c(r)$ and $c_{sr}(r)$; while the explicit form of the solution in the HNC approximation has been found to be $-c(r)/\gamma=v_L^*(r)=\mathrm{erf}(1.08x)/x$ 
\cite{ng}, 
the other solution $-c(r)$ in the SMSA has a complicated expression of the MSA type for $x<1.33$ and is simply equal to $\gamma v(r)$ for $x\geq 1.33$ where we evaluated the crossover separation $x=1.33$ from imposing the continuity of the pair correlation function
\cite{rosenfeld_lb, evans}. 
All curves of short-ranged and long-ranged potentials are cut off at longer and shorter distances, respectively; however, there is a quantitative difference between those of the LMF theory and ours though the LMF separation has been found to give better results than ours \cite{original, weeks_ocp, ng}. The deviation from the LMF division can be reduced by considering higher order terms of $\mathcal{A}\{v_L\}$ beyond the RPA, which will be presented elsewhere.

\section{Concluding remarks}

In conclusion, we have verified the applicability of the van der Waals picture to strongly coupled OCP by using the variational principle of the lower bound free energy, a first principle for the Coulomb potential separation which has been done in {\it ad hoc} ways: searching for the maximum lower bound of the free energy within the RPA valid at strong coupling, we found minus the direct correlation function as the optimal long ranged part. Since the direct correlation function has a finite value even at zero separation, the particles can overlap in the long-ranged interaction systems, one of which we have treated here field theoretically. Namely coarse-graining of particle density fields, necessary for constructing statistical field theory 
\cite{brilliantov, frusawa_lb, samuel, brown}, 
is accomplished by introducing direct correlation function as a smeared long-ranged interaction potential \cite{frusawa_lb}, about which systematic discussions such as the non-perturbative renormalization theory 
\cite{caillol} 
are to be explored. It has also been demonstrated in section 4 that the present variational theory is consistent with the conventional liquid state theory, implying that our lower bound approach applies not only to the OCP but also to any multi-scale systems including inhomogeneous counterion systems in the SC regime \cite{netz_strong, grosberg, netz, weeks_prl}.

\vspace{10pt}
\hspace{-23pt}We are grateful to K. Miyazaki for the critical reading of our manuscript. Supported by the Yamada Science Foundation and KAKENHI no. 19031027.


\end{document}